\documentstyle[12pt,aaspp]{article}
\tightenlines
\newcommand{\go}{\gtrsim}
\newcommand{\lo}{\lesssim}
\newcommand{\gam}{\Gamma}
\font\bigbf=cmbx10 scaled 1300

\begin{document}

\title{\bigbf
Hydrodynamics of Binary Coalescence.\\
 II. Polytropes with $\gam=5/3$.
}

\medskip

\author{Frederic A. Rasio\altaffilmark{1}}
\affil{Institute for Advanced Study, Olden Lane, Princeton, NJ 08540\\
       Email: rasio@guinness.ias.edu}
\and
\author{Stuart L. Shapiro\altaffilmark{2}}
\affil{Center for Radiophysics and Space Research, Cornell University,
Ithaca, NY 14853}
\altaffiltext{1}{Hubble Fellow.}
\altaffiltext{2}{Departments of Astronomy and Physics, Cornell University.}

\bigskip

\begin{abstract}

We present a new numerical study of the equilibrium and stability
properties of close binary systems.
We use the smoothed-particle hydrodynamics (SPH) technique both
to construct accurate equilibrium configurations in three dimensions
and to follow their hydrodynamic evolution.
We adopt a simple polytropic equation of state $p=K\rho^\gam$
with $\gam=5/3$ and $K=\,$constant within each star, applicable
to low-mass degenerate dwarfs as well as low-mass main-sequence stars.
For degenerate configurations, we set $K=K'$ independent of the mass
ratio.  For main-sequence stars, we adjust
$K$ and $K'$ so as to obtain a simple mass-radius relation of the form
$R/R'=M/M'$.
Along a sequence of binary equilibrium configurations for two identical stars,
we demonstrate the existence of both secular and dynamical instabilities,
confirming directly the results of recent analytic work.
We use the SPH method to calculate the nonlinear development of
the dynamical instability and to determine the final fate of the system.
We find that the two stars merge together into a single, rapidly rotating
object in just a few orbital periods.
Equilibrium sequences are also constructed for systems containing
two nonidentical stars. These sequences terminate at a Roche limit, which
we can determine very accurately using SPH.
For two low-mass main-sequence stars with mass ratio $q\lo0.4$ we find that the
(synchronized) Roche limit configuration is secularly unstable.
For $q\lo0.25$, a dynamical instability is encountered before the Roche limit.
Degenerate binary configurations remain hydrodynamically stable
all the way to the Roche limit for all mass ratios $q\ne1$.
However, unstable mass transfer can occur beyond the Roche
limit, and this is indeed observed in our numerical simulations.
Dynamically unstable mass transfer also leads to the rapid
coalescence of the binary system, although the details of the
hydrodynamic evolution are quite different.
We discuss the implications of our results for the evolution of
double white-dwarf systems and W Ursae Majoris binaries.

\end{abstract}


\section{INTRODUCTION AND MOTIVATION}

Essentially all modern studies of close binary systems are
done in the Roche approximation, where the noncompact components are modeled
as massless gas in hydrostatic equilibrium in the effective potential of
a point-mass system (see, e.g., Kopal 1978). This model applies well
to very compressible objects with centrally concentrated mass profiles, such as
giants and early-type main-sequence stars.
In contrast, most of the classical work on close binaries
was done in the completely opposite limit of a self-gravitating
 {\em incompressible\/} fluid (see Chandrasekhar 1969, and references
therein; see also Hachisu \& Eriguchi 1984b).
An essential result found in the incompressible case is that the hydrostatic
equilibrium solutions for sufficiently close binaries can become
{\em globally unstable\/} (Chandrasekhar 1975; Tassoul 1975).
Using numerical hydrodynamic calculations, Rasio \& Shapiro (1992; hereafter
RS) demonstrated that these instabilities persist in the
compressible regime, at least down to an adiabatic index as low as $\gam=2$
for two identical polytropes.

A {\em dynamically\/} unstable binary coalesces in just a few orbital periods,
forming a rapidly rotating spheroidal object surrounded by a thick disk of
shock-heated material (RS). But
well before a close binary system becomes dynamically unstable (and even if
it never does), another type of global instability can affect its evolution.
It has been referred to by various names, such as secular instability (LRS),
tidal instability (Counselman 1973; Hut 1980), gravogyro instability
(Hachisu \& Eriguchi 1984a), and Darwin instability (Levine et al.\ 1993).
Its physical origin is easy to understand (LRS1, LRS4; Rasio 1994). There
exists a {\em minimum\/} value of the total angular momentum $J$
for a {\em synchronized\/} close binary. This is simply because the spin
angular momentum, which {\em increases\/} as $r$ decreases for a synchronized
system, can become comparable to the orbital angular momentum for sufficiently
small $r$. A system that reaches the minimum of $J$ cannot evolve further
by angular momentum loss and remain synchronized. Instead, the combined
action of tidal forces and viscous dissipation will drive the system {\em
out\/} of synchronization and cause rapid orbital decay as angular momentum is
continually transferred from the orbit to the spins.

The classical analytic studies for binaries containing an
 incompressible fluid (Chandrasekhar 1969) were recently
extended to polytropes in the work of Lai, Rasio, \& Shapiro
(1993a,b, 1994a,b,c, hereafter LRS1--5 or collectively LRS).
The approach of LRS is based on the use of an energy variational
principle to construct approximate equilibrium configurations and study
their stability. The stars are modeled as self-gravitating compressible
ellipsoids, and the tidal interaction is truncated to quadrupole order.
Instabilities are identified simply from turning points appearing along
sequences of equilibrium configurations. Both secular {\em and dynamical\/}
instabilities can be identified in this way (LRS1).
Some applications to the problem of binary coalescence have
been discussed in LRS2 and LRS3. In LRS5, the method was extended to
treat the {\em dynamical\/} evolution of compressible ellipsoids.
The great usefulness of an analytic approach lies in
its simplicity. Using the method of LRS, one can
obtain an equilibrium model by solving a set of algebraic equations,
a task that can be performed on a workstation in seconds.
The dynamical evolution of binary systems can be calculated approximately
by solving ordinary differential equations (LRS5). As a result, it is
possible to explore a wide variety of simple models.
In addition, an analytic treatment can provide physical insight
into difficult issues of global stability that are easily missed
when using multidimensional numerical calculations.

There are, however, obvious limitations to an analytic treatement.
While linear stability can be investigated and orbital decay
can be tracked prior to contact, the final fate of unstable
systems remains unknown. Quantitative accuracy is limited by the
use of approximate trial functions in a variational formulation.
Stability limits obtained from a variational principle give, at best,
sufficient conditions for instability (Bardeen et al. 1977; Hunter 1977).
In the method of LRS, quantitative accuracy can be poor when the
binary configuration is close to contact or when the compressibility
of the fluid is high. In some cases, qualitatively incorrect or
unphysical results can even be obtained (see LRS4, \S5).

Coalescing binaries and stellar mergers are associated with a number
of topical problems in astrophysics.
The design of the gravitational-wave detectors for the LIGO  project
(Abramovici et al.\ 1992) is based largely on theoretical models for the
coalescence of two neutron stars. The final stage of this coalescence,
when the two neutron stars merge hydrodynamically, produces a burst of
gravitational
radiation whose characteristics probe directly the interior structure of a
neutron star.
In Paper~I (Rasio \& Shapiro 1994), we used simple polytropic models
of binary neutron stars to study the hydrodynamics of their final
coalescence and the corresponding emission of gravitational waves.
Since the nuclear equation of state at high densities is rather stiff,
we adopted high values of the adiabatic index, $\gam\go2$ (polytropic
index $n\lo1$). Here, in contrast, we consider polytropes with $\gam=5/3$
($n=3/2$). These represent fairly good models of low-mass white dwarfs
and main-sequence stars. For white dwarfs, the mass must be $\lo1\,M_\odot$,
so that the degenerate electrons are essentially nonrelativistic
(e.g., Shapiro \& Teukolsky 1983). For main-sequence stars, we must have
$M\lo0.5 \,M_\odot$, so
that most of the mass is contained in the convective envelope (cf.\ LRS4,
Table~3; Kippenhahn \& Weigert, \S 22.3).

Coalescing white-dwarf binaries are now thought to be the likely progenitors
of type Ia supernovae (Iben \& Tutukov 1984; Webbink 1984; Paczy\'nski 1985;
Yungleson et al.\ 1994). To produce a supernova,
the total mass of the system must be above the Chandrasekhar mass. Given
evolutionary considerations, this requires two C-O or O-Ne-Mg white dwarfs.
Yungelson et al.\ (1994) show that the expected merger rate for close pairs of
white dwarfs with total mass exceeding the Chandrasekhar mass is consistent
with the rate of Type Ia supernovae deduced from observations.
Alternatively, a massive enough merger may collapse to form a rapidly
rotating neutron star (Nomoto \& Iben 1985; Colgate 1990).
Chen \& Leonard (1993) have discussed the possibility that most
millisecond pulsars in globular clusters may have formed in this way.
In some cases planets may form in the disk of material ejected during
the coalescence and left in orbit around the central pulsar
(Podsiadlowski, Pringle, \& Rees 1991). Objects of planetary masses
in orbit around a millisecond pulsar (PSR B1257$+$12) have indeed been detected
(Wolszczan 1994).
A merger of two highly magnetized white dwarfs might lead to the formation of
a neutron star with extremely high magnetic field, and this scenario has been
proposed as a source of gamma-ray bursts (Usov 1992).

Close white-dwarf binaries are expected to be extremely abundant in our Galaxy.
Iben \& Tutukov (1984, 1986) predict that $\sim20$\% of all binary stars
produce close pairs of white dwarfs at the end of their stellar evolution.
The most common systems should be those containing two low-mass helium white
dwarfs.
Their final coalescence can produce an object massive enough
to start helium burning. Bailyn (1993) suggests that extreme horizontal
branch stars in globular clusters
may be such helium-burning stars formed by the coalescence of two white dwarfs.
Paczy\'nski (1990) has proposed that the peculiar X-ray pulsar 1E 2259+586
may be the product of a recent white-dwarf merger.
Planets in orbit around a massive white dwarf may also form following
a merger (Livio, Pringle, \& Saffer 1992).

Close double white dwarfs are also very important sources of low-frequency
gravitational waves
that should be easily detectable by future space-based interferometers.
Current proposals for space-based interferometers include the
LAGOS experiment (Stebbins et al.\ 1989), which should have an
extremely high sensitivity (down to an amplitude $h\sim10^{-23}$--$10^{-24}$)
to sources with frequencies in the range $\sim0.1$--$100\,$mHz.
Evans, Iben, \& Smarr (1987) estimate a white-dwarf merger rate of
order one every $5\,$yr in our own Galaxy. Coalescing systems closest to Earth
should produce quasi-periodic gravitational waves of amplitude $h\sim10^{-21}$
in the frequency range $\sim10$--$100\,$mHz.
In addition, the total number ($\sim10^4$) of close white-dwarf binaries
in our Galaxy emitting at lower frequencies $\sim0.1$--$1\,$mHz (the emission
lasting for $\sim10^2$--$10^4\,$yr before final coalescence) should
provide a continuum background signal of amplitude
$h_c\sim10^{-20}$--$10^{-21}$.
Individual sources should be detectable by LAGOS above this background when
their
frequency becomes $\go10\,$mHz.
The detection of the final burst of gravitational waves emitted during
the actual merging would provide a unique opportunity to observe in ``real
time''
the hydrodynamic interaction between the two white dwarfs, possibly followed
immediately by a supernova explosion, nuclear outburst, or some other type of
electromagnetic signal.

Main-sequence stars in the process of merging (or about to merge) are
 directly observed as W Ursae Majoris contact systems (see Rucinski 1992
for a recent review).
The question of the interior structure of these contact systems
has been the subject of a heated debate (see Shu 1980 for a summary)
and remains unresolved. At least some of the blue stragglers
observed in stellar clusters must be produced by the coalescence of two
low-mass main-sequence stars in a close binary.
Indeed, Mateo et al.\ (1990; see also Mateo 1993; Yan \& Mateo 1994)
and Kalu\'zny \& Krzeminski (1993)
have found a number of W~UMa-type contact binaries among
the blue stragglers in several globular clusters.
Large numbers of contact binaries have also been found in open
clusters (Kalu\'zny \& Rucinski 1993).
Hydrodynamic processes occurring during the final coalescence of
these binaries must play an essential role in determining the properties of
blue stragglers (Bailyn \& Pinsonneault 1994; Rasio 1993).

Our paper is organized as follows.
In \S2 we review briefly our numerical method and general conventions.
In \S3 we present our results for the equilibrium and stability
properties of binary systems containing two identical stars.
The dynamical evolution to complete coalescence is followed for
several unstable systems.
Binaries with mass ratio $q\ne1$ are treated in \S4. We consider
both the stability of equilibrium configurations up to the
Roche limit and the stability of mass transfer following Roche lobe
overflow.
The implications of our results for various types of astrophysical
systems are discussed in \S5.

\section{NUMERICAL METHOD AND CONVENTIONS}

We use the smoothed particle hydrodynamics (SPH) method for the
numerical calculations presented here.
Details about our implementation of SPH, as well as the results
of a number of test calculations and comparisons with other
numerical work are given in Paper~I and RS.
All calculations have been performed with $N\approx4\times10^4$ particles, each
particle interacting with a nearly constant number of neighbors
$N_N\approx 64$. The gravitational field is calculated in three
dimensions on a $256^3$ cartesian grid. With these resources,
equilibrium configurations can be constructed that satisfy the
virial theorem to within $\lo10^{-3}$ (cf.\ Paper~I, \S2.3 and
Fig.~1). Dynamical integrations conserve total energy to within
$\sim10^{-2}$ and angular momentum to within $\sim10^{-4}$.

Throughout this paper, numerical results are given in units
where $G=M=R=1$, where $M$ and $R$ are the mass and radius of
the {\em unperturbed\/} (spherical) primary (i.e., the more
massive of the two stars). The units of time, velocity, and
density are then
\begin{eqnarray}
t_o & = & 1600\,{\rm s}\times (R/R_\odot)^{3/2}(M/M_\odot)^{-1/2} \\
v_o & = & 440\,{\rm km}\,{\rm s}^{-1} \times
(R/R_\odot)^{-1/2}(M/M_\odot)^{1/2} \\
\rho_o & = & 6\,{\rm g}\,{\rm cm}^{-3}\times (R/R_\odot)^{-3}(M/M_\odot)
\end{eqnarray}

The mass of the secondary is denoted by $M'\le M$ and
the mass ratio is defined as $q=M'/M\le1$. For degenerate configurations,
the equilibrium radius $R'$ of the secondary is calculated assuming
constant specific entropy throughout the system, i.e., we use the
same polytropic constant $K=P/\rho^{5/3}=K'=P'/\rho'^{5/3}$ for both
components when constructing initial conditions.
This gives the equilibrium mass-radius relation $(R/R')=(M/M')^{-1/3}$.
To model binaries containing low-mass main-sequence stars, we
calculate $K'\le K$ such that the approximate mass-radius
relation $R/R'=M/M'$ is satisfied.

Particular problems are posed by the construction of accurate
equilibrium configurations and the dynamical integrations for binaries
with mass ratio $q$ far from unity (as in \S4.1 below). As in Paper~I,
we use a {\em constant number density\/} of SPH particles with unequal
masses to set up our initial conditions. Thus most of the particles
are used to model the component of larger {\em volume\/}.
This is an excellent way of maintaining good spatial resolution near the
surface of
a component about to fill its critical potential lobe.
Indeed, in a system with $q\ne1$,
it is normally the star with the larger volume that will fill its critical
lobe first (see \S4). Our method allows not only the accurate determination of
the
Roche limit along an equilibrium sequence, but also an accurate treatment
of dynamical mass transfer. This is because a reasonably large number
of SPH particles is always available near the inner Lagrangian point to feed
the accretion flow, even if the mass density there is very low.
A minor drawback of this approach is that a component with very small volume
may have to be modeled using a very small number of SPH particles.
For extreme cases, the more compact of the two stars may have to be
modeled as a single point mass (we do not treat such extreme cases in this
paper).

\section{BINARIES WITH TWO IDENTICAL COMPONENTS}

\subsection{Equilibrium and Secular Stability Properties}

Using the relaxation method at fixed binary separation $r$ described
in Paper~I (\S 2.3), we have constructed
very carefully a sequence of hydrostatic equilibrium configurations for
two identical polytropes with $\gam=5/3$.
In all cases we calculate the self-gravity of the fluid directly
in three dimensions (solving Poisson's equation by a grid-based FFT
algorithm; see RS), without making any assumptions about the density profiles
or about the smallness of the ratio $R/r$. In all cases our hydrostatic
equilibrium solutions satisfy the Virial theorem,
\begin{equation}
2T+3(\gam-1)U+W=0,
\end{equation}
where
\begin{equation}
  T=\frac{1}{2}\sum_i m_iv_i^2,~~~~~
  U=\frac{1}{\gam-1}\sum_i m_iA_i\rho_i^{\gam-1},~~~~~
  W=\frac{1}{2}\sum_i m_i\Phi_i,
\end{equation}
to better than one part in $10^3$ (cf.\ Fig.~1 of Paper~I).
Here $m_i$ and $v_i$ are the mass and velocity of
an individual SPH particle, $\rho_i$ is the local density, $\Phi_i$ is the
gravitational potential, $A_i=p_i/\rho_i^\gam$ is the entropy variable
(here equal to the polytropic constant $K$), and the
sums are over all SPH particles.

Some representative solutions along the equilibrium sequence are shown
in Figure~1. The structure of these solutions is shown both in real space
(projection
onto the orbital $x$--$y$ plane), and in terms of how the fluid fills the
effective potential wells. For a synchronized (uniformly rotating)
system with orbital frequency $\Omega_{orb}$, we calculate the effective
potential as
\begin{equation}
\Phi_e(x,y,z)=\Phi(x,y,z)-\frac{1}{2}\Omega_{orb}^2(x^2+y^2),
\end{equation}
where $x$ is along the binary axis, $z$ is along the rotation axis,
and the origin is at the center of mass of the system.
For fixed $x$, $\Phi_e$ is minimum on the binary axis ($y=z=0$),
and this minimum value is shown as a solid line in Figure~1. The fluid
fills up the space above this line, up to a level $\Phi_e^{(s)}$ which
is independent of $x$ in hydrostatic equilibrium\footnote{The value of
$\Phi_e$ for SPH particles very close to the true fluid surface is only
approximately constant, as seen in Fig.~1. This is because the {\em number
density\/} of SPH particles is not exactly constant around the surface
of a star with a large tidal deformation.}.
Along the binary axis the effective potential has a local maximum
$\Phi_e^{(i)}$ at
$x=0$ (the inner Lagrangian point) and global maxima $\Phi_e^{(o)}$
at $|x|=x_o$ (the outer Lagrangian points). There are two minima at
$|x|=x_c$, corresponding to the centers of the two components\footnote{Note
that in general $r\ne2x_c$, since we define the binary separation $r$
as the distance between the {\em centers of mass\/} of the two components.}.
Borrowing the terminology from models of W UMa binaries (Rucinski 1992 and
references therein), we define the degree of contact $\eta$ as
\begin{equation}
\eta\equiv\frac{\Phi_e^{(s)}-\Phi_e^{(i)}}{\Phi_e^{(o)}-\Phi_e^{(i)}}
\end{equation}
Clearly, we have $\eta<0$ for detached configurations and
$0<\eta<1$ for contact systems. The variation of the effective
potential and degree of contact along the sequence is illustrated in Figure~2.

It is important to realize that this equilibrium sequence
passes smoothly from detached to contact configurations as $r$ decreases.
This is in contrast to all binary equilibrium sequences
where the mass ratio $q\ne1$, which always terminate at a Roche limit before
the surfaces of the two components come into contact. For a system with
$q\ne1$, this Roche limit
corresponds to the onset of mass transfer through the inner Lagrangian
point.
For $q=1$, however, the Roche limit, which we still define as the last
equilibrium configuration along a sequence with decreasing $r$, corresponds
to the onset of {\em mass shedding through the outer Lagrangian points\/}
(cf.\ Fig.~1 for $r=2.3$).

We can use our numerical solutions to calibrate deviations from
Keplerian point-mass behavior in real, self-gravitating fluid systems.
As an example, Figure~3 shows the variation of the orbital period as
a function of $r$, compared to the Keplerian value $\Omega_K^2=
G(M+M')/r^3$. The deviation is about 1\% at first contact, and
about 6\% near the Roche limit. The values of most other equilibrium
quantities,
such as the effective radius of each component or the degree of contact
for a given binary separation, are affected at a comparable level.
These deviations imply that standard models of contact
main-sequence-star binaries based on the
Roche approximation and point-mass Keplerian relations can never be accurate
to better than a few percent. Their accuracy becomes worse for systems
in deeper contact and with lower-mass, unevolved components (which have less
centrally concentrated density profiles). In addition to these small
quantitative
changes, there are also important {\em qualitative\/} changes in the
behavior of the system, which can lead to the development
of global instabilities (LRS).

We determine the {\em secular stability limit\/} along the
equilibrium sequence by locating
the point where both the total energy $E=T+W+U$ and total angular
momentum $J$ are {\em minimum\/} (LRS). This is illustrated in Figure~4 (see
also
Fig.~2 in Paper~I, where we show a comparison with LRS and other
simpler models). Our numerical
results provide the first accurate determination of this point for a
close binary system. As can be seen in Figure~4, the minimum is
very well defined and the separate minima in $E$ and $J$ coincide to high
numerical accuracy. This is in accord with
the general property that $dE=\Omega_{orb}\,dJ$ along any
sequence of uniformly rotating fluid equilibria (Ostriker \& Gunn 1969).
Secular instability occurs very soon after contact along this sequence.
The degree of contact $\eta$ at the onset of secular instability is only
about 20\%. Thus {\em stable, long-lived equilibrium configurations can only
exist in shallow contact\/}. This result could have important implications for
the theoretical modeling of W UMa binaries (see \S5.2).

The positions and properties of various critical points along the sequence are
summarized in Table~1.

\subsection{Dynamical Stability}

One of our most significant new results is that, even for a binary
system containing a fluid as compressible as $\gam=5/3$, {\em dynamical
instabilities can still occur\/}. This is somewhat of a surprise, since
the results of LRS suggest that dynamical instabilities of close
binary equilibrium configurations can occur only for fairly stiff equations of
state,
with $\gam\go2$ (see LRS1, \S10 and Table~13; LRS3). However, the
method of LRS applies only to {\em detached\/} systems (since the two
compressible ellipsoids used to model the stars cannot overlap),
whereas the dynamical
instability found here for $\gam=5/3$ corresponds to a configuration
in deep contact, around $r=2.4$ in Figure~1. Since  tidal
effects are even stronger in contact systems than in close but
detached systems, it is reasonable to find that contact systems are even
more susceptible to global instabilities.

As in Paper~I, we determine the
dynamical stability of our equilibrium solutions by using each of them
as an initial condition for fully dynamical SPH integrations. Unstable
systems evolve to rapid coalescence in just a few orbital periods.
The results of such integrations for three of the equilibrium
solutions described in \S3.1 are illustrated in Figure~5,
where we show the time evolution of the binary separation.
The dynamical stability limit is clearly identified at $r\approx2.45$,
corresponding to a degree of contact $\eta\approx0.7$.
Each dynamical integration, using $N\approx20000$ SPH particles per star,
takes about 20 CPU hours per orbital period
on the Cornell IBM ES-9000 supercomputer.

The evolution of the dynamically unstable system with $r=2.4$ is
illustrated in Figure~6. Contours of constant density in the orbital
plane and perpendicular sections are shown at various times.
As before, we use units such that $G=M=R=1$, where $M$
and $R$ are the mass and radius of one unperturbed star.
For two main-sequence stars of mass $M\approx0.5M_\odot$ and
unperturbed radius $R\approx0.5\,R_\odot$, the initial binary separation is
$r\approx 8\times10^{10}\,{\rm cm}$ and the
initial orbital period $P_{orb}\approx3.5\,$hr.
The entire evolution shown in Figure~6 then takes about one day.
For two white dwarfs of mass $M\approx1M_\odot$ and
radius $R\approx10^{-2}\,R_\odot$, the initial binary separation is
$r\approx 2\times10^{9}\,$cm, the initial period $P_{orb}\approx25\,$s, and
the coalescence is completed in about two minutes.

We can distinguish three stages in the development of the instability.
In the linear stage, which lasts about one orbital period ($t=0$ to
$t\approx20$),
the separation between the two components decreases steadily and
the system remains close to hydrostatic equilibrium.
In the corotating frame of the binary, the relative radial velocity
remains very subsonic and the evolution remains therefore adiabatic.
When the separation has decreased below $r\approx2.3$, quasi-hydrostatic
configurations no longer exist (cf.\ Fig.~1 and~2),
and mass shedding sets in rather abruptly:  matter near the
outer Lagrangian points is ejected in a spiralling outflow
($t\approx40$--50). The evolution is still adiabatic at this stage.
In the final stage, the spiral arms widen and merge together
($t\approx50$--90).
The relative radial velocities of neighboring arms are now supersonic,
leading to some shock-heating and dissipation.
As a result, a nearly axisymmetric, rapidly rotating halo forms around
the central merged core. Essentially all of the gas in the
halo remains bound to the core (the mass loss fraction is $<10^{-3}$).
Throughout its evolution,
the system is never far from virial equilibrium (Fig.~7). This is in
complete contrast to the case of a stellar collision (see, e.g., Lai, Rasio,
\& Shapiro 1993c), where a large initial excess
of kinetic energy has to be converted into other forms.

The internal structure of the final merged configuration is similar to
that obtained and described in detail by RS for
two identical polytropes with $\gam=2$.
About 80\% of the total mass is contained inside a uniformly rotating
core where the density $\rho/\rho_c\go0.1$. The rotation inside this core is
nearly maximal, with $\Omega\simeq0.5$ in our units. About 20\% of the mass
is contained in an extended halo (outer radius $R_h\go20R$ in the equatorial
plane) of shock-heated gas in differential rotation. The structure of the
halo is pseudo-barotropic with angular velocity $\Omega\propto r_{cyl}^{-\nu}$,
where $r_{cyl}=(x^2+y^2)^{1/2}$ and $\nu\lo2$.

The ratio $T/|W|$ of kinetic energy of rotation to gravitational binding energy
(Fig.~7) is $<0.14$ at the end of the coalescence, consistent with the
axisymmetric
shape (see, e.g., Tassoul 1978). However, the final value, $T/|W|\approx0.12$,
is
considerably higher than the maximum value for stability against mass shedding
allowed for a {\em uniformly rotating\/} axisymmetric polytrope, $(T/|W|)_{max}
\approx0.06$ (see Table~3 of LRS1). This is simply because differential
rotation is very important in the halo.
Indeed, it is generally found that relaxing the assumption of uniform rotation
allows compressible equilibria with considerably higher values of $T/|W|$
to exist (see Bodenheimer \& Ostriker 1973).

%For pp:
\clearpage

\section{BINARIES WITH TWO UNEQUAL MASSES}

Equilibrium sequences with decreasing $r$ for two nonidentical stars
terminate at a Roche limit corresponding to a {\em semi-detached\/}
configuration.
Which star fills its critical lobe first depends on the mass-radius relation
for the two components. As discussed in \S2, we consider two simple cases here.
For binaries containing two degenerate stars we set $K=K'$, which leads
to the relation $(R/R')=(M/M')^{-1/3}$.
To model systems containing two low-mass main-sequence stars, we
assume $R/R'=M/M'$.

\subsection{Low-Mass Main-Sequence Stars}

With a mass-radius relation of the type $R/R'=M/M'$,
the more massive component (primary)
is also the larger in size and it has the smaller mean density of the
two. At the Roche limit, it is therefore the primary that fills its
critical lobe in the effective potential.
This is illustrated in Figure~8, where we show three configurations
with mass ratio $q=0.5$ near the Roche limit.
These configurations were constructed using the quasi-static scanning
technique described in \S2.3 of Paper~I. The minimum (Roche-limit) separation
$r_{lim}\approx2.30$--2.33 is very well determined numerically
(to better than 1\%). Binary equilibrium configurations with $r<r_{lim}$
do not exist.

If one were to continue decreasing $r$ quasi-statically below $r_{lim}$,
material from the primary would flow onto the surface of the secondary
(see Fig.~8 for $r=2.27$).
This is a possible way of continuing the equilibrium sequence beyond
the Roche limit. Unfortunately, the mass ratio is changed during this
process, becoming closer to unity, and the structure of the secondary
is no longer that of a polytrope. Since the specific entropy is lower
in the primary, the transferred material is
buoyant and remains stably on top of the lower-entropy material
in the original secondary. Eventually, when the secondary fills its
critical lobe, a contact configuration is obtained, with an interior
structure of the type considered in the contact discontinuity model of
Shu, Lubow, \& Anderson (1976). In reality, however, {\em dynamical\/}
(as opposed to quasi-static)
mass transfer would lead to shock heating of the material as it hits
the surface of the secondary (see below), and the specific entropy
in the common envelope could be considerably higher.
We do not discuss further contact systems with $q\ne1$ in this paper,
focusing instead on the properties of semi-detached configurations.

Using SPH, we have constructed several equilibrium sequences for decreasing
values of the mass ratio.
In each case we determine the secular stability of the Roche
limit configuration by following the variation of the total angular
momentum $J$ along the sequence\footnote{Total energy can be considered
also, and leads to the same conclusions. However, by comparing our results
to those of LRS for large $r$, we find that our numerical
accuracy in the calculated value of $J$ is generally better.}.
The results are shown in Figure~9. We see that {\em there exists a critical
value of the mass ratio $q_{crit}\approx0.4$ where the Roche limit
configuration becomes secularly unstable\/}. For $q<q_{crit}$, there exists
an {\em unstable branch\/} of configurations between the minimum of $J$ and the
Roche limit. Possible implications of these results for Algol and W~UMa
binaries are discussed in \S5.1 below.

Figure~9 also shows a comparison of our SPH results with those obtained in
LRS4 using a semi-analytic variational method.
In all cases, the equilibrium $J(r)$ curves determined by the two
methods are in excellent quantitative agreement (to within a few
parts in $10^3$) up to
to the Roche limit as determined by SPH. However, the method of LRS
cannot provide the {\em location\/} of the Roche limit to better than about
20\% in $r$. A similar situation was encountered in LRS1 (cf. \S 3 of that
paper)
 with respect to the mass-shedding limit along equilibrium sequences
of isolated, rapidly rotating stars.
The variational method of LRS can be used to determine equilibrium
properties of uniformly rotating polytropes quite accurately up to the mass
shedding limit, but it cannot by itself predict accurately
the position of the mass shedding limit.

Noting that the Roche limit configuration for $q=0.25$ is located far beyond
the secular stability limit along the sequence, we decided to test its
{\em dynamical\/} stability. Recall that the onset of dynamical instability
 always occurs further along the sequence (at smaller $r$) than the
the secular stability limit. However, secularly unstable configurations
may remain dynamically stable all the way to the Roche limit (see LRS4,
\S3.3). For polytropes with $n=1.5$ and $R/R'=M/M'$, Figure~4 of LRS4
does in fact predict that configurations with sufficiently small mass
ratio should become dynamically unstable.

The results of several dynamical SPH integrations for binaries
with $q=0.25$ are illustrated in Figure~10. There is indeed a
dynamical instability developing just before the Roche limit along
the equilibrium sequence, at $r=r_{dyn}\approx2.05$--$2.1
>r_{lim}\approx2.01$--2.03.
For an initial configuration with $r=2.05$, we find that the orbit
decays rapidly, leading almost immediately to the onset of dynamical mass
transfer onto the small secondary (Fig.~10c).
Numerical integrations starting from $r=2.1$ and $r=2.2$ were also
performed and showed only small epicyclic oscillations (as in Fig.~5),
with no sign of coalescence. Thus the dynamical stability limit
for the sequence with $q=0.25$ is at $r_{dyn}\approx2.05$--2.1.
Sequences with even smaller values of $q$ may be susceptible
to the same dynamical instability well before reaching the Roche limit.
However, we caution that the interpretation of our numerical results for
this case is complicated by the highly unstable mass transfer following
Roche lobe overflow. Indeed, the mass transfer is from the more massive
to the less massive component, implying a tendency for the separation
to decrease during mass transfer, in addition to the adiabatic increase
in radius of the mass-losing star (see, e.g., Shore, Livio, \& van den Heuvel
1994).
Since the numerically determined dynamical stability limit is so close to the
Roche limit, we cannot strictly rule out the possibility
 that the instability we see developing
is in fact driven mostly, or even entirely, by the unstable response
of the system to the mass transfer\footnote{The statement
made in \S5 of LRS4 that all sequences with
$0.25\le p<1$ are dynamically stable may therefore be correct. However,
it should have read $0.25< p<1$. (The mass ratio is denoted by $p$ in LRS)}.
Since our simple models for main-sequence stars can apply
only to a fairly small range of masses, $0.1\lo M/M_\odot \lo 0.5$,
the question of what really happens to these models for extreme mass
ratios is obviously of limited astrophysical interest.

\subsection{Degenerate Configurations}

For degenerate systems with $q\ne1$, the more massive (and denser) component
typically suffers very little tidal deformation compared to the other,
less massive component (cf.\ LRS4, \S4.4). As a result, the overall
destabilizing effect of tidal interactions in the binary system is
greatly reduced. Using SPH to construct equilibrium sequences, we
find that, for all mass ratios, {\em close binaries containing two degenerate
stars remain secularly (and, therefore, dynamically) stable all the
way to the Roche limit\/}.
An example is shown in Figure~11 for $q=0.5$. The equilibrium configuration
with $r=3.9$ is very close to the Roche limit and its angular
momentum is still well above the minimum value $J_{min}\simeq0.86$
calculated using the method of LRS4.

Stability of all hydrostatic equilibrium configurations does not
say anything about how the system will respond to mass transfer after
the loss of angular momentum drives it beyond the Roche limit.
Since the mass-losing component responds adiabatically by expanding
in size, one may expect the mass transfer itself to be dynamically
unstable. Unless the orbit can expand sufficiently to counteract
the tendency of the mass-losing star to overflow its Roche lobe even more,
the mass-transfer rate will grow catastrophically (Shore et al.\ 1994).

We demonstrate this instability of the mass transfer itself by
studying the dynamical evolution of the (stable) equilibrium configuration
shown in Figure~11a. This configuration is as close to the
Roche limit as we could realize numerically, with just a few SPH particles
having crossed the inner Lagrangian point. The initial separation is $r=3.90$
and the corresponding orbital period is $P_{orb}\approx39$ in
our units. The evolution is shown in Figures~12 and~13.
We see that the mass transfer is indeed unstable, with the
mass transfer rate increasing on a timescale comparable to the orbital
period. As before (cf.\ Figs.~5 and~10b), we have checked the stability
of systems with slightly larger separation, $r=4.0$ and $r=4.1$,
by calculating their dynamical evolution for several orbital periods,
and found no sign of instability.
The mass transfer rate increases very slowly at first, with the secondary's
mass loss being only 10\% after about 4 complete orbital periods (Fig.~13a),
but then accelerates rather abruptly.
Complete tidal disruption of the secondary is observed after about
five orbital periods (Fig.~12e--g). The final configuration is again an
axisymmetric
merger with a core-halo structure, the core containing the primary
and the halo being made of shock-heated material from the
disrupted secondary. Note the pronounced cusp in the isodensity contours
near the equatorial plane (Fig.~12h), indicating a maximal rotation rate.
To construct Figure~13, we have determined the structure of the
``instantaneous''
effective potential (eq.\ [6]) at every time during the evolution.
This requires iterations, since the instantaneous value of $\Omega_{orb}$
is determined from the motion of the
fluid contained inside the (inner) critical potential lobe.
The orbital angular momentum $J_{orb}$ corresponds to the
center-of-mass motion of the two components. Clearly, orbital angular
momentum is {\em lost\/} during mass transfer, leading to rapid orbital decay
and the unstable increase in the mass transfer rate (Fig.~13a).

The evolution of the density and specific entropy profiles along the
binary axis is illustrated in Figure 13b. The density inside the primary
{\em decreases\/} during the evolution. This is because shock-heating as
the massive accretion column impacts the surface (Fig.~12b--d)
leads to a slight increase in the specific entropy throughout the interior
of the primary. The material that accumulates on top of the primary's surface
inside its critical potential lobe is of very low density (high entropy),
and its mass is essentially negligible in determining the final structure
of the core (in the absence of shock-heating, the slight increase in mass
of the primary would make its density {\em increase\/}). Outside of the
critical lobe, rotational support is important. However, the final merged
configuration has $T/|W|\approx0.08$, far below the secular stability limit
for axisymmetric configurations at $T/|W|\approx0.14$ (see, e.g., LRS1).

As we mentioned in the introduction, coalescing white-dwarf binaries are
potentially important sources of low-frequency gravitational waves.
In Figure~14, we show examples of gravitational radiation waveforms
emitted during the final coalescence, both for a dynamically unstable
system with $q=1$ (\S3.2), and for the system with $q=0.5$ driven
by unstable mass transfer (considered above). These were calculated
as in Paper~I and RS, in the quadrupole approximation. The two waveforms
are strikingly different. One obvious difference in overall timescale
comes from our normalization to the mass and radius of the primary.
But there are also important {\em qualitative differences\/}
caused by the different hydrodynamic processes at work.
For the system with $q=0.5$ driven by unstable mass transfer, the amplitude
of the waves {\em decreases\/} while the frequency remains essentially constant
(it actually decreases also, but very slightly).
For the dynamically unstable system with $q=1$, both the amplitude and
the frequency {\em increase\/}.
In both cases, however, the emission shuts off abruptly at the very
end of the evolution. This is because the system evolves rapidly to a
stationary axisymmetric configuration, which does not radiate gravitational
waves (cf.\ Paper~I; RS).

\section{DISCUSSION}

\subsection{Evolution of W UMa Binaries}

It is tempting to envision (cf.\ Rasio 1993, 1994)
that the secular orbital decay of a contact system
through loss of angular momentum (by gravitational radiation
or via magnetized winds), probably accelerated by viscous dissipation
(see below), could eventually bring the two components
sufficiently close together to make the system dynamically unstable.
The following hydrodynamic evolution leads to the rapid coalescence
of the two stars into a single, rapidly rotating object (Fig.~6i,j)
which one could call a ``proto-blue-straggler''.

It should be stressed, however, that hydrodynamic calculations can only
reveal the new hydrostatic equilibrium configuration reached by the system
 following complete coalescence. This configuration is in {\em mechanical\/}
equilibrium, but far from thermodynamic equilibrium. On a timescale much longer
than the hydrodynamic timescale, various dissipative transport processes will
be
operating that can modify considerably the structure of the
object. Thus the  results of purely
hydrodynamic calculations cannot directly predict the {\em observable
properties\/} of blue stragglers. For example, angular momentum may well
be removed very efficiently from the rapidly and differentially rotating
outer halo through a magnetized wind (cf.\ Leonard \& Clement 1993).
The outer layers of the observed blue straggler could then appear
to be rotating slowly, even if the interior is still spinning rapidly.
Thus the rapid rotation predicted by hydrodynamic calculations (which
conserve angular momentum) may never be
 observed directly. Indeed, at least some blue stragglers are known
observationally to be slow rotators (Mathys 1991).

Our results do not confirm the suggestion by Williams \& Roxburgh (1976)
that {\em all\/} contact systems with $q=1$ should be dynamically unstable
to mass transfer from one component to the other. This was their proposed
explanation for why W UMa systems with $q$ close to unity are not observed.
It should be noted that such a system may now in fact have been found
(Samec, Su, \& Dewitt 1993).
We have calculated the dynamical evolution of a number of contact
configurations with $q=1$ (cf.\ Fig.\ 5). We have not detected any tendency for
an asymmetric unstable mode to develop, at least within a few orbital periods.
Even during the evolution of a
dynamically unstable system to coalescence (Fig.~6), there is no sign
of an asymmetry developing. It is important to note that no assumption
was made about a mirror symmetry between the two components in our
calculations.
Each star was modeled separately using a completely independent set of SPH
particles and the gravitational field is calculated on a cartesian
grid without any assumption about the geometry.
In addition, our polytropic models respond adiabatically by expanding when they
lose mass, since $(\partial R/\partial M)_K<0$. This
should make a binary system particularly susceptible to
mass-transfer instabilities. Instead, the model of a $0.6\,M_\odot$
main-sequence star considered in detail by Williams \& Roxburgh (1976)
has $(\partial R/\partial M)_S>0$.
Strictly speaking, however we cannot rule out that an {\em overstable\/}
asymmetric mode of oscillation may exist for all contact configurations with
$q=1$. This mode could grow in amplitude over a timescale long compared
to the few orbital periods that we have followed to establish the {\em
orbital\/} stability of the equilibrium solutions.
Indeed, the dynamical instability
of the orbit that we identified numerically in RS and analytically in LRS
corresponds to a truly {\em unstable\/} (as opposed to overstable) mode of
oscillation of the binary system. In the presence of this unstable mode,
the separation $r$ will never increase again once it starts  decreasing
(cf.\ Fig.~5). Thus a dynamical integration following just one complete
epicyclic oscillation is in principle sufficient to establish whether a
particular
solution is unstable.

Our results concerning the {\em secular stability\/} of contact binaries have
important implications for the interpretation of their observed
characteristics.
To our knowledge, secular (tidal) instabilities have never been discussed
before
in the context of W UMa binaries, although van't Veer (1979) correctly
identified
the minimum of $J$ (in a simple binary model containing two rigid spheres;
cf.\ LRS4, \S6.3) as a critical point in the evolution of contact systems.
The orbital decay of a secularly unstable binary
 proceeds on the (de)synchronization timescale,
which is typically much shorter than the timescale associated with
any angular-momentum loss mechanism (see LRS4, \S6).
For two low-mass main-sequence stars with deep convective
envelopes, this timescale could be as short as $10^4\,$yr (Zahn 1977).
This may explain why the components of W~UMa binaries are always observed
to be in very shallow contact (Rucinski 1992).
If the two components ever got closer together because of mass
exchange or loss of angular momentum, the system would become secularly
unstable and the two stars would quickly coalesce.
The results of \S3 suggest that the degree of contact at the onset of
secular instability is indeed small, at least for low-mass stars with
mass ratio not too far from unity.
The exact location of the secular stability limit (and the corresponding
maximum value of the degree of contact for a stable, long-lived system) should
depend sensitively on the internal structure of the binary, and, in
particular, the distribution of specific entropy in the system.
Thus a large sample of well-determined contact-binary parameters could
be used to place interesting constraints on their internal structure.

We have not explored the stability of mass transfer from one main-sequence
star to another in this paper. This is in part because the simple polytropic
model
that we have adopted is not realistic enough. All our polytropic models
would clearly lead to dynamically unstable mass transfer, independent of the
mass
ratio. In reality, mass transfer from one main-sequence star to another
may be dynamically stable or unstable depending on the details
of the internal structure of each component and the mass ratio.
We did, however, present a complete calculation of dynamically unstable
mass transfer for a {\em degenerate\/} system (\S4.2), mainly to
illustrate the differences between a dynamical coalescence driven by unstable
mass
transfer and one driven by a true dynamical instability of the binary
equilibrium configuration. For two low-mass main-sequence stars, both
instabilities can develop simultaneously, as demonstrated in \S 4.1.
The results of \S4.1 suggest that even when the mass
transfer is dynamically stable, the orbital evolution of the binary
may be affected by secular instabilities for sufficiently small mass ratio.
If the Roche limit configuration is secularly unstable,
mass transfer could be driven entirely by viscous dissipation, on a
timescale much shorter than that of the angular momentum loss.

\subsection{Coalescence of Double White Dwarf Systems}

Our results show that
hydrodynamics can play an important role in the coalescence of two
white dwarfs, either because of dynamical instabilities of the
equilibrium configuration (\S3), or following the onset of dynamically unstable
mass transfer (\S4.2). Here also, a more detailed study of the stability
of mass transfer between two white dwarfs in general would require the
use of a more realistic equation of state.
A calculation of dynamical mass transfer between two higher-mass white dwarfs
was performed by Benz et al.\ (1990) using a
more realistic degenerate equation of state plus thermal gas pressure.
Although the results of Benz et al.\ are qualitatively similar to ours,
one may question
their use of a nonequilibrium initial configuration. They started their
calculation with two {\em spherical\/} stars, and with the secondary
overflowing its critical lobe by almost 10\% in radius.
Not too surprisingly, they found that the secondary was then tidally disrupted
in less than an orbital period. In contrast, starting the calculation
from a true equilibrium configuration (Figs.~11a and~12a) leads to a much
more gradual increase in the mass transfer rate, and complete disruption
of the secondary occurs only after several orbital periods during which the
evolution is quasi-static.

For two massive enough white dwarfs, the merger product may be well above the
Chandrasekhar mass $M_{Ch}$.
It may therefore explode as a (Type Ia) supernova, or perhaps collapse
to a neutron star.
The rapid rotation (cf.\ Fig.~6) and possibly high mass (up to $2M_{Ch}$)
of the object must be taken into account for determining its final fate.
This is not done in current theoretical calculations of accretion
induced collapse (AIC), which always consider a
nonrotating white dwarf just below the Chandrasekhar limit
accreting matter slowly and quasi-spherically (Canal et al.\ 1990;
Nomoto \& Kondo 1991; Isern 1994).
Under these assumptions it is found that collapse to a neutron star
is possible only for a narrow range of initial conditions.
In most cases, a (Type Ia) supernova explosion follows the ignition of the
nuclear fuel in the degenerate core.
However, the fate of a much more massive object with substantial
rotational support formed by dynamical coalescence may be very different.

\acknowledgments

This work has been supported by a Hubble Fellowship to F.~A.~R.\ funded by NASA
through Grant HF-1037.01-92A from the Space Telescope Science Institute,
which is operated by AURA, Inc., for NASA, under contract NAS5-26555.
Partial support was also provided by NSF Grant AST 91--19475 and
NASA Grant NAGW--2364 to Cornell University. F.~A.~R.\ also acknowledges
the hospitality of the ITP at UC Santa Barbara.
This research was conducted using the resources of the Cornell Theory
Center, which receives major funding from the NSF and IBM Corporation,
with additional support from the New York State Science and Technology
Foundation and members of the Corporate Research Institute.

\clearpage

\begin{table}

\caption{CRITICAL POINTS ALONG THE EQUILIBRIUM SEQUENCE
          FOR TWO IDENTICAL STARS\tablenotemark{a} }

\bigskip

\begin{tabular}{l c c c c c}
Property:             &  $r$   &  $\eta$  &   $P_{orb}$   &   $E$     &     $J$
  \\
\tableline

First Contact         &  2.77  &  0.      &    20.2       &  -1.0142  &  1.335
  \\

Secular Instability   &  2.67  &  0.22    &    19.0       &  -1.0153  &  1.333
  \\

Dynamical Instability &  2.45  &  0.7     &    16.5       &  -1.0062  &  1.358
  \\

Roche Limit\tablenotemark{b}
                      &  2.35  &  1.      &    15.2       &  -1.0013  &  1.374
  \\

\end{tabular}

\tablenotetext{a}{Units are defined in \S 2.2; $r$ is the binary separation,
$\eta$ is the degree of contact (eq.\ [7]), $P_{orb}$ the orbital period,
and $E$ and $J$ are the total equilibrium energy and angular momentum.}

\tablenotetext{b}{Defined as the equilibrium configuration with the
minimum binary separation.}

\end{table}

\clearpage

%FIG 1
\begin{figure}
\caption{Sequence of binary equilibrium configurations for two identical
polytropes with $\gam=5/3$. Projections of all SPH particles onto the
orbital plane are shown in the upper portion of the figure. Projections onto
the
$(x,\Phi_e)$ plane, where $x$ is the coordinate along the binary axis
and $\Phi_e$ is the effective potential (eq.~[6]), are shown at the bottom.
The solid lines show the variation of $\Phi_e(x,y=0,z=0)$ along the binary
axis. Units are such that $G=M=R=1$ as defined in \S 2. Contact configurations
are obtained
when the binary separation $r\lo2.8$. For $r\lo2.4$, mass shedding through the
outer Lagrangian points occurs, and no equilibrium configuration
exists.
}
\end{figure}

%FIG 2
\begin{figure}
\caption{
Variation of the critical potentials along the equilibrium sequence
shown in Fig.~1. Values of the effective potential $\Phi_e$
at the outer Lagrangian points (o), the inner Lagrangian point (i), the
fluid surface (s), and at the center of each star (c) are shown as a
function of binary separation (above). The degree of contact $\eta$
(eq.~[7]) is shown below. The dashed lines show the positions
of the first contact configuration ($\eta=0$) and the Roche limit
($\eta=1$).
}
\end{figure}

%FIG 3
\begin{figure}
\caption{
Variation of the orbital period (in the unit of eq.~[1]) along the equilibrium
sequence shown in Fig.~1. The dashed line shows the value of the Keplerian
orbital period for two point masses with the same separation $r$.
The strong tidal interaction between the two stars makes the effective
interaction potential stronger than $1/r$, and shortens the rotation period
compared to a point-mass system.
}
\end{figure}

%FIG 4
\begin{figure}
\caption{
Variation of the total energy and angular momentum along the equilibrium
sequence shown in Fig.~1. The vertical dashed line shows the position of
the secular stability limit, where both $E$ and $J$ are minimum.
}
\end{figure}

%FIG 5
\begin{figure}
\caption{
Time evolution of the binary separation $r$ computed from dynamical
calculations
starting from several of the equilibrium models shown in Fig.~1. Equilibrium
configurations with $r\lo2.45$ are dynamically unstable.
}
\end{figure}

%FIG 6
\begin{figure}
\caption{
Evolution of the unstable binary with $r=2.4$ shown in Figs.~1 and~5.
Contours of equal density in the orbital plane are shown as a function
of time.
The scale is logarithmic, with 16 contours covering 4 decades down from
the value $\rho_{max}\simeq1.5$ in the unit of eq.~[3]. The orbital
rotation is counterclockwise. Vertical sections are shown in~(e) and~(j).
All other plots show sections through the orbital plane.
}
\end{figure}

%FIG 7
\begin{figure}
\caption{
Evolution of the Virial ratio $(2T+2U+W)/|W|$ (cf.\ eq.~[4]) and
the ratio $T/|W|$ of kinetic energy of rotation to gravitational binding
energy during the dynamical coalescence shown in Fig.~6.
}
\end{figure}

%FIG 8
\begin{figure}
\caption{
Equilibrium configurations near the Roche limit for a binary containing
two low-mass main-sequence stars with mass ratio $q=0.5$. Conventions are
as in Fig.~1. The more massive (less dense) star is on the left. The
Roche limit is clearly identified at a binary separation $r=r_{lim}
\approx2.30$--$2.33$.
}
\end{figure}

%FIG 9
\begin{figure}
\caption{
Variation of the total angular momentum $J$ as a function of binary
separation $r$ along several equilibrium sequences for two low-mass
main-sequence stars. The solid lines show our numerical results,
terminating at the Roche limit configuration along each sequence
(indicated by a solid vertical segment). The dashed lines are
from the semi-analytic calculations of LRS4. When the mass ratio
$q\lo0.4$, there exists a branch of secularly unstable configurations
between the minimum of $J$ and the Roche limit.
}
\end{figure}

%FIG 10
\begin{figure}
\caption{
Dynamical evolution of a system containing two low-mass main-sequence
stars with mass ratio $q=0.25$.
In (a), we show the equilibrium configuration with $r=2.05$, just below the
Roche
limit; conventions are as in Figs.~1 and~8.
The more massive but less dense star (primary) is on the left.
In (b), we show the dynamical evolution of the separation $r$ for systems
with three different initial separations: the system with $r=2.05$, shown
in (a), is clearly unstable, while those with $r=2.1$ and $r=2.2$ are stable.
The horizontal dashed line marks the position of the Roche limit (onset of
mass transfer).
In (c) we show how the unstable binary with $r=2.05$ has evolved when $t=25$.
For comparison, in (d), we also show the stable system with $r=2.1$ at the same
time $t=25$. In (c) and (d), conventions are as in Fig.~6.
}
\end{figure}

%FIG 11
\begin{figure}
\caption{
Roche limit configuration for a system containing two low-mass
white dwarfs with $q=0.5$. In (a), conventions are as in Figs.~1 and~8.
The lower-mass (and lower-density) star is on the right, about to
start mass transfer onto the higher-mass star on the left.
In (b), conventions are as in Fig.~9.
}
\end{figure}

%FIG 12
\begin{figure}
\caption{
Dynamical evolution of a system containing two low-mass white dwarfs
with $q=0.5$.
The initial binary separation is $r=3.9$, just beyond the Roche limit.
Although all equilibrium configurations down to the Roche limit are
stable, the mass transfer itself is dynamically unstable, leading to
the complete coalescence of the binary in about five orbital periods
($P_{orb}\approx39$ at $t=0$). Conventions are as in Fig.~6.
}
\end{figure}

%FIG 13
\begin{figure}
\caption{
Evolution of various quantities during dynamically unstable
mass transfer for the binary system shown in Fig.~12.
In (a) we show the separation $r$, mass of the secondary $M'$, and
orbital angular momentum $J_{orb}$ as a function of time.
In (b), we show the density and specific entropy ($s\propto\log[p/\rho^\gam]$)
profiles along the binary axis at t=0 (solid lines), t=100 (long-dashed
lines), t=200 (short-dashed lines), and $t=220$ (dotted lines).
}
\end{figure}

%FIG 14
\begin{figure}
\caption{
Gravitational wave amplitude $h_{+}$ as a function of retarded time
for the dynamical evolution shown in Fig.~12.
Here quantities are labeled in geometrized units ($G=c=1$).
The amplitude is shown for an observer situated at a distance $r_O$
along the rotation axis. For comparison, we also show the amplitude of
the waves emitted during the dynamical coalescence of a system with $q=1$.
}
\end{figure}

\end{document}